\documentclass[preprint2]{aastex}

\newcommand{\ergcms}{ergs cm$^{-2}$ s$^{-1}$}

\newcommand{\ergss}{ergs s$^{-1}$}

\newcommand{\cpspcu}{counts s$^{-1}$ PCU$^{-1}$}

\begin{document}
\title{XTE J1946+274 = GRO J1944+26: An Enigmatic Be/X-ray Binary}
\author{Colleen. A. Wilson\altaffilmark{1}, Mark H. Finger\altaffilmark{2}}
\affil{SD 50 Space Science Research Center, National Space Science and 
Technology Center, 320 Sparkman Drive, Huntsville, AL 35805}
\email{colleen.wilson-hodge@msfc.nasa.gov}
\author{M.J. Coe}
\affil{Dept. of Physics and Astronomy, The University, Southampton, SO17 1BJ,
England}
\author{Ignacio Negueruela\altaffilmark{3}}
\affil{Observatoire de Strasbourg, 11 rue de l'Universit\'e, 67000 Strasbourg, France}
\altaffiltext{1}{NASA's Marshall Space Flight Center}
\altaffiltext{2}{Universities Space Research Association}
\altaffiltext{3}{Now at: Dpto. de F\'{\i}sica, Ingenier\'{\i}a de Sistemas y 
Teor\'{\i}a de la Se\~{n}al, Universidad de Alicante, Apdo. de Correos 99, 
E03080, Alicante, Spain}
\begin{abstract}
XTE J1946+274 = GRO J1944+26 is a 15.8 s Be/X-ray pulsar discovered simultaneously
in 1998 September with the Burst and Transient Source Experiment (BATSE) on the
{\em Compton Gamma Ray Observatory (CGRO)} and the All-Sky Monitor (ASM) on the
{\em Rossi X-ray Timing Explorer (RXTE)}. Here we present new results from BATSE
and {\em RXTE} including a pulse timing analysis, spectral analysis, and 
evidence for an accretion disk. Our pulse timing analysis yielded an orbital 
period of 169.2 days, a moderate eccentricity of $0.33$, and implied a mass 
function of 9.7 M$_{\odot}$. We observed evidence for an accretion disk, a 
correlation between measured spin-up rate and flux, which was fitted to obtain a
distance estimate of $9.5 \pm 2.9$ kpc. XTE J1946+274 remained active from 1998 
September - 2001 July, undergoing 13 outbursts that were not locked in orbital 
phase. Comparing {\em RXTE} PCA observations from the initial bright outburst in
1998 and the last pair of outbursts in 2001, we found energy and intensity 
dependent pulse profile variations in both outbursts and hardening spectra with
increasing intensity during the fainter 2001 outbursts. In 2001 July, optical 
H$\alpha$ observations indicate a density perturbation appeared in the Be disk 
as the X-ray outbursts ceased. We propose that the equatorial plane of the Be 
star is inclined with respect to the orbital plane in this system and that this
inclination may be a factor in the unusual outburst behavior of the system.
\end{abstract}

\section{Introduction}
Be/X-ray binaries are the most common type of accreting X-ray pulsar systems. 
They consist of a pulsar and a Be (or Oe) star, a main sequence star of spectral
type B (or O) that shows Balmer emission lines (See e.g., Slettebak 1988 and
Coe 2000 for reviews.)\nocite{Coe00,Slettebak88} The line emission is believed to be associated with 
an equatorial outflow of material expelled from the rapidly rotating Be star 
which probably forms a quasi-Keplerian disk near the Be star \citep{Hanuschik96,Quirrenbach97}.
X-ray outbursts are produced when the pulsar interacts with this disk. Be/X-ray
binaries typically show two types of outburst behavior: (a) giant outbursts (or
type II), characterized by high luminosities ($L_{\rm X} \gtrsim 10^{37}$
\ergss) and high spin-up rates  (i.e., a 
significant increase in pulse frequency) and (b) normal outbursts (or type I), 
characterized by lower luminosities ($L_{\rm X} \sim 10^{36}-10^{37}$ \ergss), 
low spin-up rates (if any), and recurrence
at the orbital period \citep{Stella86,Bildsten97}. As a population Be/X-ray 
binaries show a correlation between their spin and orbital periods
\citep{Corbet86,Waters89}. 

On 1998 September 8 a 15.8 s pulsar, which was designated GRO J1944+26, was
discovered with the Burst and Transient Source Experiment \citep[BATSE]{Fishman89}
on the {\em Compton Gamma Ray Observatory (CGRO)}. GRO J1944+26 was localized to
a $5\arcdeg \times 8\arcdeg$ error box \citep{Wilson98}. At the same time, the 
All-Sky Monitor \citep[ASM]{Levine96} on the {\em Rossi X-ray Timing Explorer 
(RXTE)} discovered a new source which they localized to a $6\arcmin\ \times 
26\arcmin$ error box and designated XTE J1946+274 \citep{Smith98}. Subsequent 
observations with the Proportional Counter Array \citep[PCA]{Jahoda96} also on 
the {\em RXTE} on 1998 September 16 revealed 15.8 s pulsations, confirming that
BATSE and {\em RXTE} were seeing the same object \citep{Smith98}. Scanning 
observations with the PCA further improved the position to a 2.4\arcmin\ error 
circle \citep{Takeshima98}. Observations with {\em BeppoSAX} further improved 
the X-ray position to R.A. = $19h45m38s$, Decl. = +27\arcdeg 21\arcmin.5 
(equinox 2000.0; error radius 1\arcmin\ at 95\% confidence) \citep{Campana98}. 
XTE J1946+274 lies in the error box of the Ariel 5 transient 3A 1942+274; 
however, there is a non-negligible probability that this is a chance association
\citep{Campana99}.  

A likely optical counterpart has been identified as an optically faint 
$B \sim 18.6$ mag, bright infrared ($H \sim 12.1$) Be star
at R.A. = $19h45m39s$, Decl. = 27\arcdeg22\arcmin 00\arcsec, which lies in 
the refined BeppoSAX error circle\citep{Verrecchia02}. Spectroscopic and
photometric data indicate a B0-1IV-Ve star at 8-10 kpc. The counterpart's mass 
is not well constrained, but is expected to be in the $10-16 M_{\odot}$ range. 
Two previously proposed candidates \citep{Israel98, Ghavamian98} lie 
$10\arcsec$\ outside the {\em BeppoSAX} error circle.

Using data from the 1998 outburst taken with the PCA and the High Energy X-ray 
Timing Experiment \citep[HEXTE]{Rothschild98} on {\em RXTE}, \citet{Heindl02} revealed a cyclotron
resonance scattering feature or cyclotron line near 35 keV. This feature
implies a magnetic field strength of $3.1(1+z) \times 10^{12}$ G, where $z$ is
the gravitational redshift of the emission region. Using about 1 year of {\em RXTE} ASM
2-10 keV flux measurements, beginning with the 1998 outburst, \citet{Campana99}
observed 5 outbursts in which they found evidence of an $\sim 80$ day
modulation in the outburst flux. They found that the first outburst in September
1998 was markedly different in rise and decay timescales than the 4 following
outbursts and based on this suggested that the first outburst might be
classified as a giant outburst.  XTE J1946+274 was observed with the Indian 
X-ray Astronomy Experiment \citep[IXAE]{Paul01} 1999 September 18-30 (MJD
51408-51421) and 2000 June 28-July 7 (MJD 51723-51733). They detected 15.8 s pulsations in the 2-6 and 6-18 keV bands,
with similar double peaked profiles in both observations. The pulse period
history was consistent with a constant intrinsic spin-up plus an eccentric 
orbit. The reported pulse periods were $P = 15.78801(04)$ s and on MJD 
51445.0 and $P = 15.76796(18)$ s on MJD 51727.5.

XTE J1946+274 was active from 1998 September through 2001 July, when it
dropped below the detection threshold of the {\em RXTE} PCA. As of 2002 July, no
additional outbursts have been seen with the {\em RXTE} ASM. Table~\ref{tab:obs}
lists the dates and instruments observing XTE J1946+274's location. In this paper we 
will present results from observations of XTE J1946+274 with BATSE and {\em
RXTE}, including histories of pulse frequency and total flux for 
XTE J1946+274. From the pulse frequency history, we derive an orbital solution. 
Applying this orbital solution, we will investigate the observed correlation 
between spin-up torques and observed flux and discuss implications for the 
accretion mechanism. We will describe the unusual orbital phasing of the 
outbursts. Using data from the {\em RXTE} PCA and HEXTE, we will investigate 
spectral variations in the initial bright outburst and the last 2 outbursts. We
compare our X-ray results to optical H$\alpha$ observations. 

\section{Analysis and Results}
\subsection{Frequency Search}
To determine pulse frequencies for XTE J1946+274, we performed
a grid search over a range of candidate frequencies using data from the Large 
Area Detectors on BATSE\citep{Fishman89}. This search technique is described
in detail in \citet{Finger99} and \citet{WilsonHodge99}. We will briefly
summarize the technique here. Variations of this technique, including searches 
in pulse period, have been widely used \citep[for example]{Buccheri83, Bildsten97}.

The pulsar search technique consists of 3 steps (1) data selection and 
combination, (2) 20-50 keV pulse profile estimates, and (3) a grid search in 
frequency. First, the BATSE DISCLA channel 1 (20-50 keV, 1 s time resolution) 
data were selected for which the source was visible, the high voltage was on, 
the spacecraft was outside the South Atlantic Anomaly, and no electron 
precipitation events or other anomalies were flagged by the BATSE mission 
operations team. The count rates were combined over the 4 LADs viewing XTE 
J1946+274, using weights optimized for an exponential energy spectrum, $f(E) = 
A \exp(-E/kT)$ with temperature $kT = 12$ keV, and grouped into $\approx 300$ s 
segments. A segment length of 300 s was used because it was short enough that 
the background was well-fitted by a quadratic model. The segment boundaries were
chosen to avoid Earth occultation steps from bright 
sources\citep {WilsonHodge99}. 

The second step in this process was estimation of an initial 20-50 keV pulse 
profile for each segment of data. In each segment, the combined rates were 
fitted with a model consisting of a sixth-order Fourier expansion in pulse phase
(representing the 20-50 keV pulse profile) 
$r(t) = \sum_{h=1}^6 \alpha_{hk} \exp (i 2 \pi h \phi_0(t))$
where $\alpha_{hk}$ is the estimated complex Fourier 
coefficient for harmonic\footnote{In this paper, harmonics are defined as $n\nu$
where $n = 1,2,3, \ldots$ and $\nu$ is the pulse frequency.} $h$ in segment $k$
and a spline function with quadratics in time (representing the background plus
mean source count rate). A sixth-order Fourier expansion was chosen based on the number of harmonics that
are significant in a 1-day observation, the typical integration time required to
detect XTE J1946+274. Our 
initial pulse phase model was of the form $\phi_0(t) = \nu_0 (t-t_0)$, where 
$\nu_0 = 63.1513$ mHz before MJD 51270 and $\nu_0 = 63.2800$ mHz after MJD 
51270 were barycentric frequencies and $t$ was the time corrected to the Solar 
System barycenter\footnote{In this paper, the words barycentered and barycentric
refer to this correction.} using the JPL DE-200 ephemeris \citep{Standish92}. 
The initial pulse frequency was changed after MJD 51270 because continued 
spin-up of the pulsar moved the observed frequency too far away from the
folding frequency. The value and slope of the spline function were required to 
be continuous across adjacent 300 s segment boundaries, but not across data 
gaps. 

Ideally we would like to be able to fit a pulse profile to the entire 4-day
interval of data for each frequency grid point; however this is computationally very
very expensive. Instead we have formulated an equivalent method in which we 
first fit pulse profiles to short segments of data using a single fixed 
frequency. Then the profiles are combined and shifted by a frequency offset
corresponding to each grid point. Fitting pulse profiles to short 
data segments has other advantages in addition to improving computational efficiency. (1) If 
the folding frequency chosen was incorrect, the profile will be smeared out. 
However, the degree of smearing in pulse phase $\Delta \phi$ is approximately 
given by $\Delta \phi = \Delta \nu L_{\rm s} n$, where $\Delta \nu$ is an offset
in frequency, $L_{\rm s}$ is the segment length, and $n$ is the number of 
Fourier coefficients. For $L_{\rm s} = 300$ s, $\Delta \nu = 0.17$ mHz, and 
$n = 6$, $\Delta \phi = 0.31$, which corresponds to a drop in the amplitude of 
the sixth harmonic by 15\%. The amplitude of the first harmonic only drops by 
about 0.4\%. Using a short segment reduces smearing of the pulse profile. (2) 
Typically 470 pulse profiles from the 300 s segments are accumulated within each
4-day interval. Since we have a large number of profiles we can estimate sample
variances for each Fourier coefficient. Sample variances give us a measure of 
the noise present in the data and account for aperiodic noise if it is present.  

The final step in our advanced pulsar search was a grid search in frequency
using the set of typically several hundred estimated 20-50 keV pulse profiles 
from 4-day intervals of data. New XTE J1946+274 pulse frequencies were 
determined from an initial grid search over 1200 evenly spaced trial barycentric 
frequencies in the range 62.97-63.32 mHz for MJD 48361-51270 and 63.10-63.45
mHz for MJD 51270-51690. The search range was changed because continued spin-up
of the pulsar moved the pulse frequencies outside the original search range. 
For each frequency offset $\Delta \nu$, a mean pulse profile is computed for
each 4-day interval. The mean pulse profile is of the form 
\begin{equation}
\bar r(t) = \sum_{h=1}^n \bar \alpha_h \exp ( i 2 \pi h \phi(t))
\end{equation}
where $\phi(t) = \phi_0(t) + \Delta \phi_k$, $\Delta \phi_k = 
\Delta \nu (t_k-t_0)$, and $t_k$ is the midpoint barycentric time of segment
$k$. The mean coefficients $\bar \alpha_h$ are given by
\begin{equation}
\bar \alpha_h = \sum_{k=1}^M w_k \alpha_{hk} \exp (-i 2 \pi h \Delta \phi_k)
\end{equation} 
with weights $w_k = \sigma_{\alpha_{hk}}^{-2} ( \sum_{\ell=1}^M
\sigma_{\alpha_{h\ell}}^{-2})^{-1}$.

The typical statistic used for a search in pulse frequency or pulse period is 
the $Z_n^2$ statistic \citep{Buccheri83} given by 
\begin{equation}
Z_n^2 = \sum_{h=1}^n \frac{|\bar \alpha_h|^2}{\sigma_{\bar \alpha_h}^2} 
\end{equation}
where $\sigma_{\bar \alpha_h}$ is the formal (i.e., Poisson statistical) error 
on the Fourier coefficient $\bar \alpha_h$, respectively.  If aperiodic noise is
present, either due to XTE J1946+274 or due to other sources in the large BATSE
field of view, e.g., Cygnus X-1, the Poisson statistical error is an 
underestimate because the actual underlying noise level is higher than Poisson 
noise. This can create a problem with the $Z_n^2$ statistic, causing it to 
depend on the noise level. To account for aperiodic noise, we used all of the 
typically $\sim 470$ pulse profiles from 300 s segments within each 4-day interval to 
estimate sample variances for the Fourier coefficients for each harmonic in the
mean profile. Each harmonic was treated separately. We then modified the
$Z_n^2$ statistic by replacing the Poisson variances with our computed sample 
variances to create a new statistic which we will call $Y_n$ after 
\citet{Finger99} where it is described in detail. Due to the large 
field-of-view of BATSE, other pulsars were often also present when we were 
measuring XTE J1946+274. If we limited our statistic to use the first 3 
harmonics where XTE J1946+274 was the brightest, we reduced the chances of 
contamination of the search results from other pulsars that happen to have 
harmonics near the higher harmonics of XTE J1946+274. The best-fit frequency for
each 4-day interval was then determined using the $Y_3$ statistic. 

Using a similar method we also generated pulse frequency measurements for 
{\em RXTE} PCA observations during the initial outburst in 1998
September-October and during the last 2 outbursts, 2001 March-July. Barycentered
Standard 1 (125 ms, no energy resolution) data were fitted with a model 
consisting of a constant background plus a 6th-order Fourier expansion in 
pulse phase model, creating an estimated 2-60 keV pulse profile for each PCA 
observation. The pulsed phase model consisted of a constant barycentric 
frequency, $\nu_0 = 63.17466$ mHz for the 1998 September-October observations and
$\nu_0 = 63.4239$ mHz for the 2001 observations, estimated from projections of
BATSE measurements. For each set of observations, we searched over a grid of 151
evenly spaced frequencies spanning the range $\nu_0 \pm 0.17$ mHz, the same size
interval as used for BATSE. 

\begin{figure}
\plotone{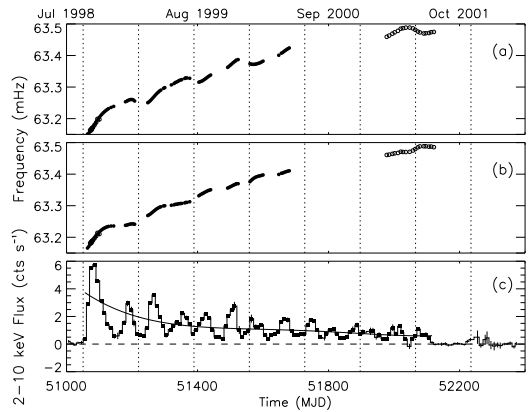}
\caption{XTE J1946+274 outburst history. (a) Barycentered pulse frequency
measurements with BATSE (filled circles) and {\em RXTE} PCA (open circles).
(b) The pulse frequency measurements shown in (a) have been corrected for our 
orbital solution listed in the fourth column of Table~\protect\ref{tab:orb} (c) 2-10 keV flux history
measured with the {\em RXTE} ASM. Each point is a 10-day average flux. Filled
squares denote 3-$\sigma$ or better detections. The solid curve denotes a
quartic polynomial fitted to the {\em RXTE} ASM flux (see Section 2.3).
Vertical dotted lines in all three panels denote the estimated times of 
periastron passage using the orbital parameters in the fourth column of
Table~\protect\ref{tab:orb}.
\label{fig:freqs}}
\end{figure}

Figures~\ref{fig:freqs}a \& c show the barycentered pulse frequency history and
the 2-10 keV total flux history, respectively, for XTE J1946+274. In 
Figure~\ref{fig:freqs}a, the trend in the pulse frequencies repeatedly changes 
from spin-up to spin down and back to spin-up at regular intervals (every ~170 
days). This repeated pattern is most likely due to the pulsar's orbit. Within 
each 170 day cycle in the pulse frequencies, there are two outbursts.
 
\subsection{Orbit Fitting}

\begin{figure}
\plotone{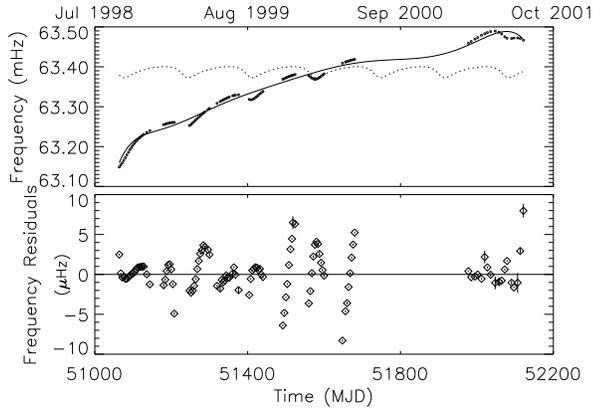}
\caption{Top: The solid curve denotes the emitted frequency modeled as a 10th 
order polynomial. The dotted curve illustrates the magnitude of orbital effects.
The open circles denote the full pulse frequency model including both the 10th
order polynomial and the orbital effects. Bottom: Pulse frequency residuals for
the model consisting of a global orbit and a global 10th order emitted frequency
model.
\label{fig:poly}}

\end{figure}

Examination of the observed pulse frequency history showed us two things: (1)
a strong orbital signature and (2) strong intrinsic torque variations. To
attempt to extract orbital parameters, we first fitted the observed pulse 
frequencies $\nu_{\rm obs}$ with a model consisting of a global orbit and a 
global polynomial frequency model, i.e.,
\begin{equation}
\nu_{\rm obs} = \nu_{\rm emit}(1-\beta)\label{eqn:fobs}
\end{equation}
where $\nu_{\rm emit} =  \sum_{k=0}^n a_k (t-t_0)/k!$ is a model of the emitted 
frequency. The velocity relative to the observer  $\beta=v/c$ is given by
\begin{equation}
\beta = \frac{2 \pi a_{\rm X} \sin i}{P_{\rm orb} c}
\left\{\frac{(1-e^2)^{1/2} \cos \omega \cos E - \sin \omega \sin E}
 {(1-e \cos E)} \right\}
\end{equation}
where $a_{\rm X} \sin i$ is the projected semi-major axis of the pulsar's orbit; 
$P_{\rm orb}$ is the orbital period; $\omega$ is the periapse angle; and $e$ is
the orbital eccentricity; $E$ is the eccentric anomaly, given by 
\begin{equation} 
E-e \sin E = 2 \pi \frac{(t-\tau_{\rm p})}{P_{\rm orb}} 
\end{equation}
and $\tau_{\rm p}$ is the epoch of periastron passage.  We fitted models
containing polynomials of orders 1-10. All of these models were very poor fits.
Table~\ref{tab:orb} lists the orbital parameters obtained using a first and 10th
order polynomial for comparison. Errors on these parameters have been inflated 
by a factor of $(\chi^2/\rm{d.o.f.})^{1/2}$, but are still expected to be considerably 
underestimated because the $\chi^2$ values are so large. 
Figure~\ref{fig:poly} shows the emitted frequency model $\nu_{\rm emit}$,
$a_0(1-\beta)$ the constant term from the emitted frequency model times the
velocity relative to the observer to illustrate the amplitude of orbital
effects, the full frequency model including the emitted frequency and orbit, and
the frequency residuals for the 10th order polynomial model.

\begin{figure}
\plotone{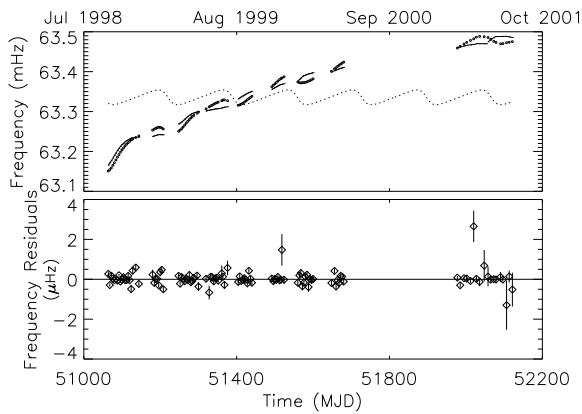}
\caption{Top: The broken curve denotes the piecewise emitted frequency model. 
The dotted curve illustrates the magnitude of orbital effects. The open circles
denote the full pulse frequency model including both the piecewise emitted
frequency model and the orbital effects. Bottom: Pulse frequency residuals for
the model consisting of a global orbit and piecewise emitted frequency
model.
\label{fig:resid}}
\end{figure}

Clearly a simple model such as a global polynomial did not suffice to describe 
the intrinsic torques. Instead we decided to model the intrinsic torques in a
piecewise fashion. Each outburst was split into 2-6 intervals, depending on the
length of the outburst, where each interval contained 3-4 frequency 
measurements and spanned 16-24 days or 16-32 days for the BATSE and the {\em
RXTE} PCA data, respectively. In each interval, we fitted the observed pulse 
frequencies with a global orbit plus an independent linear 
frequency model, Equation~\ref{eqn:fobs} with $\nu_{\rm emit} = \nu_i+\dot \nu_i
(t-t_i)$ as a model of the emitted frequency for each time interval $i$. 
Table~\ref{tab:orb} lists our best fit orbit. Figure~\ref{fig:resid} shows
$\nu_{\rm emit}$, $\nu_{16}(1-\beta)$ to illustrate the amplitude of orbital
variations (Note $\nu_{16}$ was chosen because it lies in near the center of the
frequency range.), the full frequency model including the emitted frequency and
orbit, and the frequency fit residuals. This fit is not formally 
acceptable either; however, we believe this is mostly because this model still 
does not completely describe the intrinsic torques. Errors on the individual 
parameters in Table~\ref{tab:orb} have been inflated by 
$(\chi^2/{\rm d.o.f.})^{1/2}$. Comparing the three columns of Table~\ref{tab:orb}
shows that the orbital parameters appear to be similar, whether we use a simple
polynomial model or a more complicated piecewise model for $\nu_{\rm emit}$. 
Figure~\ref{fig:freqs}b shows the orbit-corrected spin frequencies. The regular
pattern seen in Figure~\ref{fig:freqs}a has been removed, while intrinsic torque
effects remain in the data. 

\subsection{Orbital Phasing of Outbursts\label{sec:orbph}}

A close look at Figure~\ref{fig:freqs} indicates that the outbursts are not
fixed in orbital phase and that we are seeing approximately 2 outbursts per
orbit. To study the orbital phase of the outbursts, we fitted the 2-10 keV {\em
RXTE} ASM flux in Figure~\ref{fig:freqs}c with a quartic polynomial, and used
this polynomial to normalize the outbursts. We then computed the orbital
position using
\begin{eqnarray}
x = a (\cos E - e) \\
y = a (1-e^2)^{1/2} \sin E 
\end{eqnarray}
where $a$ is the semi-major axis, estimated from $a_{\rm X} \sin i$ assuming
$i = 70\arcdeg$ which gives a companion mass of $\sim 14.5 M_{\odot}$. 
Figure~\ref{fig:orbphase} shows the normalized intensity versus orbital 
position, with darker colors indicating higher intensities. The outer loop of
the spiral has $a = 323 R_{\odot}$, the estimated radius assuming $i =
70\arcdeg$. For each successive orbit, we reduced $a$ by 10\% to offset the 
orbits. This plot spirals in solely to allow comparison of outburst phases. We 
are not claiming that the neutron star is spiraling in toward its companion. 

The outburst peaks clearly are not fixed in orbital phase, although there is
typically one peak on each side of the orbit. Just before periastron and just
before apastron, the flux is typically relatively low. The cause of these low
flux periods is unclear. Although the fluxes show considerable modulation, they
remain detectable with {\em RXTE} throughout the period 1998 September - 2001 
July.

\subsection{Spectral Analysis\label{sec:spec}}

\begin{figure}
\plotone{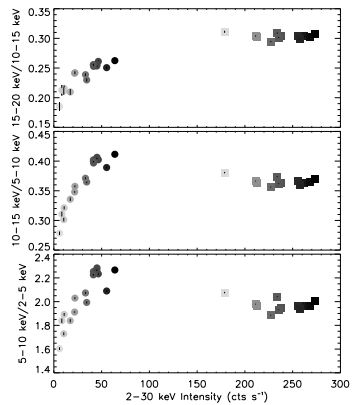}
\caption{Hardness-intensity diagrams. Filled squares denote points from the
1998 outburst and filled circles denote the 2001 outbursts. The color darkens as
the intensity increases within each outburst to allow comparison with
Figure~\protect\ref{fig:cc}
\label{fig:HI}}
\end{figure}

\begin{figure}
\plotone{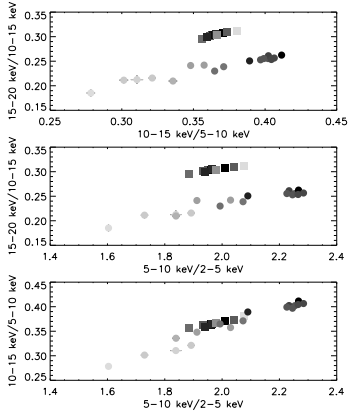}
\caption{Color-color diagrams. Filled squares denote points from the
1998 outburst and filled circles denote the 2001 outbursts. For both sets of
outbursts, the color darkens as the intensity increases.
\label{fig:cc}}
\end{figure}

To look for variations in energy spectra during the outbursts observed
with the {\em RXTE} PCA, we first studied hardness ratios. Using
FTOOLS\footnote{\url{http://heasarc.gsfc.nasa.gov/ftools}} 5.1
\citep{Blackburn95} and Standard2
(16-s, 129 energy channel) data, we created light curves in 4 energy bands 
corresponding to 2-5, 5-10, 10-15, and 15-20 keV for both the 1998 and 
2001 outbursts.  These light curves were background subtracted and their times
were corrected to the Solar System barycenter using FTOOLS. Next we formed 3 
hardness ratios from adjacent energy bands. Figure~\ref{fig:HI} shows each of 
these ratios vs. total PCA source count rate in the 2-30 keV band. 
Figure~\ref{fig:cc} shows color-color diagrams. In both figures, the grey-scale
denotes intensity, with darker points denoting higher intensities. (Note: This
is the measured 2-30 keV intensity, not the normalized 2-10 keV intensity used 
in Section~\ref{sec:orbph} and Figure~\ref{fig:orbphase}.) For all of the hardness ratios, we see a correlation between
hardness and intensity in the 2001 outburst that is not present in the 1998
outburst. In the color-color diagrams, the hardness ratios are correlated for
both sets of outbursts and have similar slopes. However, during the 2001
outbursts, points move from left to right along the correlation as intensity
increases; while in the 1998 outburst, the points move around in hardness as
intensity increases.

\begin{figure}
\plotone{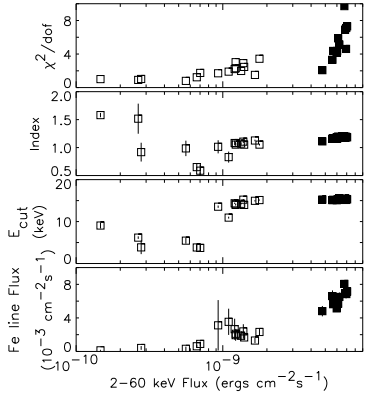}
\caption{Spectral fitting results for an absorbed power law with a high energy
cutoff and a Gaussian iron line. Shown are the goodness of fit $\chi^2/$d.o.f.,
the power law photon index, the cut off energy $E_{\rm cut}$, and the integrated
flux in the iron line versus 2-60 keV flux. Filled squares denote the 1998
outburst and open squares denote the last pair of outbursts observed in 2001.
\label{fig:highecut}}
\end{figure}

\begin{figure}
\plotone{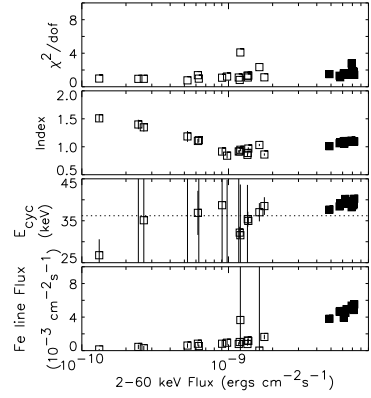}
\caption{Spectral fitting results for an absorbed power law with a cyclotron
absorption feature and a Gaussian iron line. Shown are the goodness of fit 
$\chi^2/$d.o.f., the power law photon index, the cyclotron line energy $E_{\rm
cyc}$, and the integrated flux in the iron line versus 2-60 keV flux. The dotted
line denotes the cyclotron energy measured during the 1998 outburst by
\citet{Heindl02}. Filled squares denote the 1998 outburst and open squares 
denote the last pair of outbursts observed in 2001.
\label{fig:cyclabs}}
\end{figure}

To better quantify the observed spectral changes, we generated response
matrices, count spectra, and background spectra using FTOOLS for PCA Standard2 
data and HEXTE event mode data for each observation. We then fitted each of the
observations in XSPEC\footnote{\url{http://xspec.gsfc.nasa.gov}} 11.1 \citep{Arnaud96}
with two models: (1) an absorbed power law with a high-energy 
cutoff and an iron line, XSPEC model: PHABS (POWERLAW + GAUSSIAN) HIGHECUT$^7$
and (2) an absorbed power law with an iron line and a cyclotron absorption 
feature, XSPEC model: PHABS (POWERLAW + GAUSSIAN) CYCLABS$^7$. For both models 
we used only data in the 2.7-30 keV energy range for PCA data and in the 
15-50 keV range for HEXTE. The power law normalizations were allowed to be 
independent for the PCA and each HEXTE cluster. Figure~\ref{fig:highecut} shows
$\chi^2$, spectral index, cutoff energy, folding energy, and flux in the Fe line
vs. 2-60 keV flux. Figure~\ref{fig:cyclabs} shows $\chi^2$, spectral index, the 
cyclotron line energy, and the flux in the Fe line vs. 2-60 keV flux. Other 
spectral parameters did not show significant variations with intensity.

A comparison of the two models showed that model 2 generally provided a better
fit to the data (see Figures~\ref{fig:highecut} \& \ref{fig:cyclabs}, top
panel). In both models, the power law photon index remains relatively
independent of flux for fluxes above about $10^{-9}$ \ergcms, however, below
that flux level, the power law index softens as the intensity decreases. The 
integrated flux in the iron line is correlated with the 2-60 keV flux in both
models, indicating that the iron line is indeed a feature of XTE J1946+274's
spectrum and not a background feature. In model 2 a cyclotron absorption
feature, when it is constrained, tends to be near the value observed by
\citet{Heindl02}. 

To estimate an upper limit to the quiescent flux from XTE J1946+274, we fitted 
PCA data from each of 6 observations after pulsations were undetectable in 2001
August - September with an absorbed power law. These observations were 
consistent with a power law index of $\sim 2.4$, a hydrogen column of about 
$6 \times 10^{22}$ cm$^{-2}$, and a 2-30 keV flux of $(4-7) \times 10^{-12}$
\ergcms. Since {\em RXTE} is not an imaging instrument, we cannot distinguish
between low-level emission from XTE J1946+274 and background sources such as the
galactic ridge. However, this gives us an upper limit on the flux from XTE
J1946+274.  

\subsection{Spin-up Torque-Flux correlations\label{sec:fdotvsflux}}

\begin{figure}
\plotone{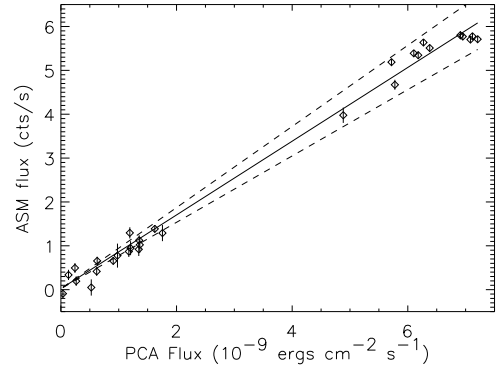}
\caption{ASM count rate vs. PCA 2-60 keV flux measurements. The solid line
denotes our best fit the the correlation. This fit was used to estimate a
bolometric correction to the ASM data. Dashed lines correspond to our best
fit slope $\pm 10$\% indicating our estimate of systematic error due to
changing beaming angles at high fluxes. 
\label{fig:bc}}
\end{figure}

\begin{figure}
\plotone{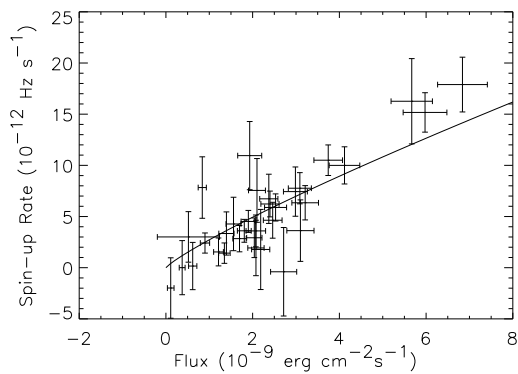}
\caption{Spin-up rate vs. bolometrically corrected ASM flux. The solid line is
our best fit of the simple torque model with a distance of 9.5 kpc.
\label{fig:fdotvsflux}}
\end{figure}

To estimate a bolometric correction to the {\em RXTE} ASM measurements, we 
plotted the 2-60 keV fluxes from model 2 in Section~\ref{sec:spec} vs. the 5-day
average ASM count rate surrounding each PCA observation. We then fitted a line 
to these data, obtaining a bolometric correction of 
1 ASM ct s$^{-1} = (1.2 \pm 0.1) \times 10^{-9}$ \ergcms (2-60 keV). The
error on this correction has been estimated including systematic effects.
Figure~\ref{fig:bc} shows the best fit line to these data. Near the peak flux
level, the ASM fluxes show evidence for a turnover relative to the PCA fluxes.
We have attempted to account for this turn-over by including a systematic error
of 10\% on our bolometric correction.
Using this bolometric correction, we then plotted the average bolometrically
corrected ASM flux vs. spin-up rate for each interval in our frequency model. 
Figure~\ref{fig:fdotvsflux} clearly shows a correlation between spin-up rate
and flux. Such a correlation suggests a disk is present because direct wind 
accretion is believed to be less efficient at transferring angular momentum 
\citep{Ruffert97, Benensohn97, Ruffert99}. If enough angular momentum is 
present, a disk will form.

Simple accretion theory assumes that the material from the companion star is
flowing onto a rotating neutron star with a strong magnetic field. This field
determines the motion of material in a region of space called the
magnetosphere. The size of this region is denoted by the magnetospheric radius
$r_m$, given by \citep{Pringle72, Lamb73}
\begin{equation}
r_{\rm m} \simeq k (G M_{\rm X})^{-1/7} \mu^{4/7}  \dot M^{-2/7} \label{eqn:rm} 
\end{equation}
where $G$ is the gravitational constant; $M_{\rm X}$ is the mass of
the neutron star; and $\dot M$ is the mass accretion rate, which is assumed to 
be related to the observed bolometric flux $F$ by $\dot M = 4 \pi d^2 F R_{\rm 
X} / G M_{\rm X}$, where $R_{\rm X}$ is the radius of the neutron star and $d$ 
is the distance to the pulsar.  $k$ is a constant factor of order 1. 
Equation~\ref{eqn:rm} with $k \simeq 0.91$ gives the Alfv\'en radius  $r_A$ for
spherical accretion and with $k \simeq 0.47\ n(\omega_s)$ gives the 
magnetospheric radius derived by \citet{Ghosh79}. 

The torque applied by 
accretion of matter onto a neutron star, assuming torques due to matter leaving
the system are negligible, is given by \citep{Lamb73} 
\begin{equation}
\frac{d}{dt}(2 \pi I \nu) = \dot M \ell \label{eqn:torque}
\end{equation}
where $I$ is the moment of inertia of the neutron star and $\ell$ is the
specific angular momentum of the material. If $I$ is assumed constant, then
$\ell$ is given by
\begin{equation} 
\ell = 2 \pi I \dot \nu \dot M^{-1}, \label{eqn:ell}
\end{equation}
where $\dot \nu$ is the spin-up rate. To estimate $\ell$ for XTE J1946+274, we 
assumed typical pulsar parameters, $M_{\rm X} = 1.4 M_\odot$, $R_{\rm X} = 10$ 
km, $I = 10^{45}$ g cm$^2$, typical values of $\dot \nu \simeq 5 \times 
10^{-12}$ Hz s$^{-1}$ and $F_{\rm bol} \simeq 2 \times 10^{-9}$ \ergcms (see 
Figure~\ref{fig:fdotvsflux}), and a distance $d = 8-10$ kpc 
\citep{Verrecchia02}. This yielded $\ell = (2.4-3.8) \times 10^{17}$ cm$^{2}$
s$^{-1}$.  

An accretion disk will form if the specific angular momentum of 
the material accreted from the Be star's disk is comparable to the Keplerian 
specific angular momentum at the magnetospheric radius, i.e., 
\begin{equation}
\ell \simeq \ell_{\rm m} = (G M_{\rm X} r_m)^{1/2}. \label{eqn:ellrm}
\end{equation}
For XTE J1946+274, using the magnetic field measurement $B=3.1
(1+z) \times 10^{12}$ G \citep{Heindl02} which implies $\mu = 1.6 \times 
10^{30}$ G cm$^3$, then $\ell_{\rm m} = (2.7-2.8) \times 10^{17}$ cm$^{2}$ 
s$^{-1}$ for $k=0.91$ and $\ell_{\rm m} = (1.9-2.0) \times 10^{17}$ cm$^{2}$ 
s$^{-1}$ for $k=0.47$. For XTE J1946+274, $\ell = 0.9-1.3 \ell_{\rm m}$ for 
$k=0.91$ and $\ell = 1.3-1.9\ell_{\rm m}$ for k = 0.47, hence an accretion disk
is most likely present. In 
contrast, for the wind-fed system Vela X-1 where a disk is not expected to be 
present, $\dot \nu \simeq 6 \times 10^{-14}$ Hz s$^{-1}$ \citep{Inam00}, $L 
\simeq 2 \times 10^{38}$ \ergss, and $\mu \simeq 2.1 \times 10^{30}$ G cm$^{3}$
\citep{Makishima99} leading to $\ell \simeq 3.5 \times 10^{14}$ cm$^2$ s$^{-1}$
and $\ell_{\rm m} \simeq 2.2 \times 10^{17}$ cm$^2$ s$^{-1}$, i.e., $\ell 
\simeq 0.002 \ell_{\rm m}$. Further, three-dimensional simulations of wind
accretion show that the average specific angular momentum accreted via wind
accretion is always smaller than the Keplerian value \citep{Ruffert97,
Ruffert99}.

Since an accretion disk appears to be present, we can use simple accretion
theory to derive a distance to XTE J1946+274.
Substituting Equations~\ref{eqn:ellrm} into Equation~\ref{eqn:torque} and 
solving for $\dot \nu$ gives
\begin{equation}
\dot \nu = 1.39 k^{1/2} (G M_{\rm X})^{-3/7} R_{\rm X}^{6/7} I^{-1} 
\mu^{2/7} d^{12/7} F^{6/7}. 
\end{equation} 
Assuming $k=0.91$ and the parameters defined earlier,
\begin{equation}
\dot \nu_{12} = 5.8 \times 10^{-2} d_{\rm kpc}^{12/7} F_9^{6/7},
\label{eqn:nudot}
\end{equation}
where $\dot \nu_{12}$ is the spin-up rate in units of $10^{-12}$ Hz s$^{-1}$ and
$F_9$ is the bolometric flux in units of $10^{-9}$ \ergcms. 
We then fitted this model to the data in Figure~\ref{fig:fdotvsflux}. To account
for the flux and $\dot \nu$ errors, we computed $\chi^2$ as
\begin{equation}
\chi^2 = \sum_{i=1}^n \left\{\frac{(F_i-F^{\rm mod}_i)^2}{\sigma_{F_i}^2} + 
\frac{(\dot\nu_i - \dot \nu^{\rm mod}(F_i^{\rm mod}))^2}{\sigma_{\dot \nu_i}^2}
\right\}
\end{equation}
where $F_i^{\rm mod}$ is a free parameter fitted to each flux and 
$\dot \nu^{\rm mod}(F_i^{\rm mod})$ is Equation~\ref{eqn:nudot} evaluated at 
$F_i^{\rm mod}$. The solid line shows our best fit, with $\chi^2/32 = 1.0$, 
which gives us a distance of $9.5 \pm 0.3$ kpc. This distance is consistent with
optical observations of the counterpart. The errors on the distance are computed
from the fit itself and take into account only the statistical errors on the
flux and spin-up rate and the estimated systematic error on the bolometric
correction. However, there is considerable uncertainty in our assumed neutron
star parameters. The estimated distance depends on the following combination of
neutron star parameters: $M_x^{1/4} R_x^{-1/2} I^{7/12}$. Using the neutron star
equations of state given by \citet{Wiringa88} and assuming $M_x = 1.0-1.8
M_{\odot}$, we estimate the error on the distance due to uncertainty in the
neutron star parameters is about 30\%, i.e., $d = 9.5 \pm 2.9$ kpc.

\subsection{Pulse Profiles}

Using FTOOLS and {\em RXTE} PCA event mode data, we created light curves and 
corresponding background light curves in 5 energy bands: 2-5, 5-10, 10-15, 
15-20, and 20-30 keV for each of 29 observations, 12 from the 1998 outburst and 
17 from the 2001 outbursts. Each light curve was barycentered, background
subtracted, and folded at the appropriate frequency from our frequency
search. Each pulse profile contained 50 phase bins. First we examined profiles 
from the entire 2-30 keV band to search for intensity dependent variations. We 
aligned all of the pulse profiles by finding the phase of the minimum of the 
profile using a quadratic interpolation and placing that minimum at phase 0.0. 
We found definite variations with intensity in both the 1998 and 2001 outbursts
and also evidence that the pulse profile was different at similar intensities 
during the rise and fall of the initial bright outburst. 

\begin{figure}
\epsscale{1.5}
\plottwo{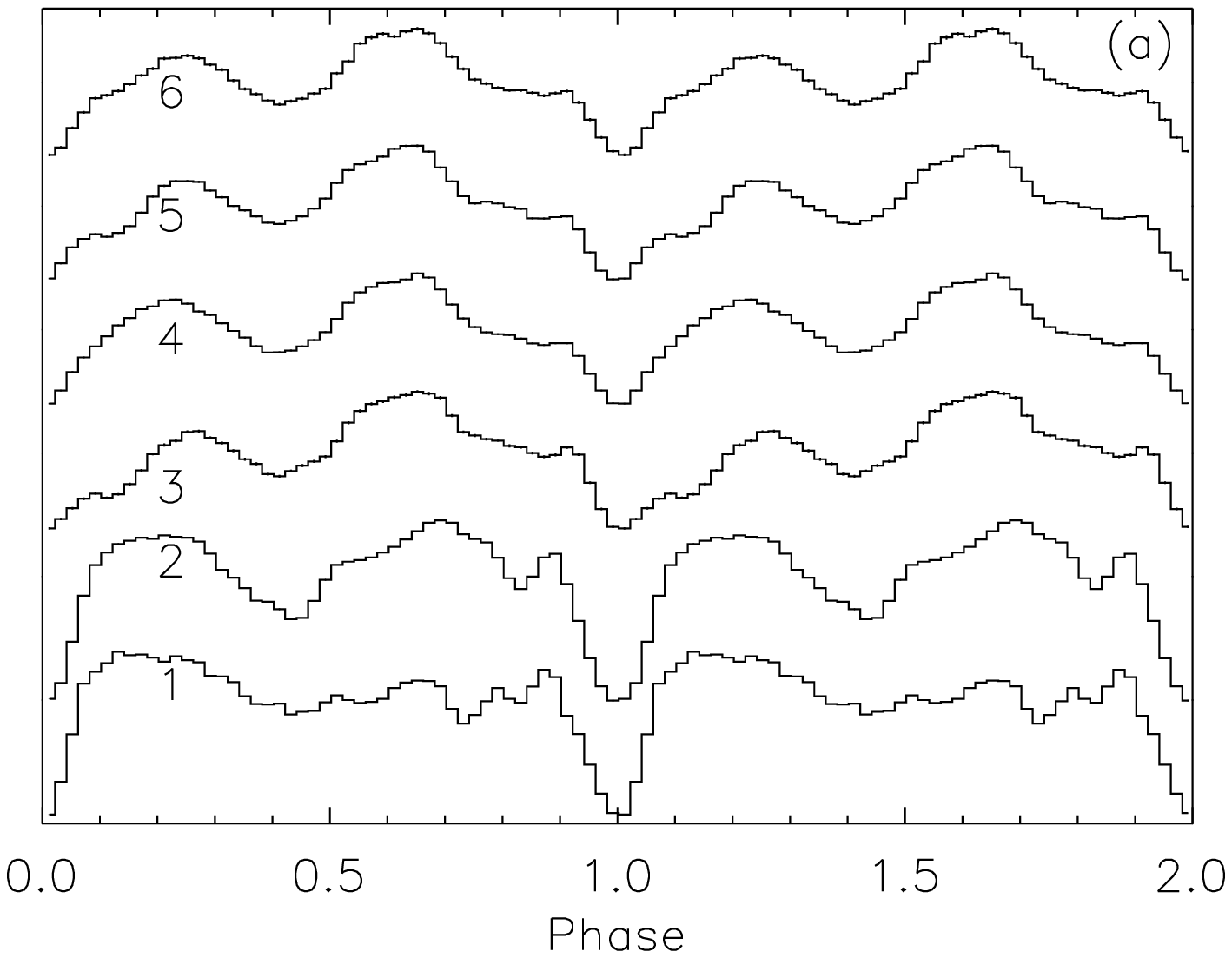}{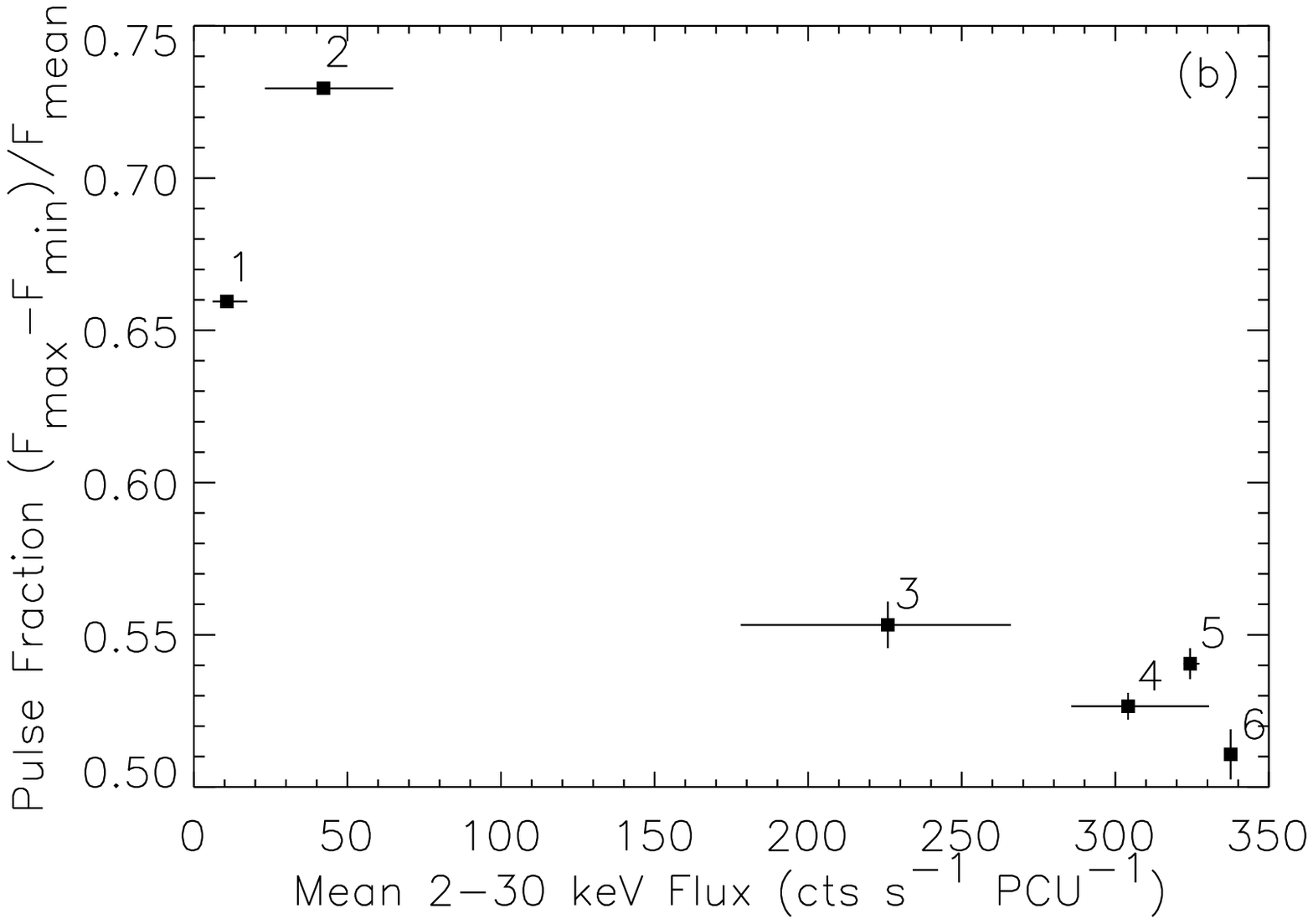}
\caption{(a): Average 2-30 keV pulse profiles from 6 intensity bands versus 
intensity. Profiles are numbered 1-6, corresponding to descriptions in the main
text. Profiles 1 and 2 are from the 2001 outbursts. Profiles 4 and 5 are from
similar intensities during the rise and fall, respectively, of the 1998 
outburst. Each profile has been normalized to its mean flux. Profiles are offset
by arbitrary amounts to keep them from overlapping. (b): Peak-to-peak pulse
fraction versus mean 2-30 keV flux for the pulse profiles shown in panel (a). 
Horizontal bars indicate the flux range spanned by the profiles included in the
average.
\label{fig:intprof}}
\epsscale{1.0}
\end{figure}

Profiles with similar shapes and similar intensities were averaged to better
illustrate the observed shape changes. Figure~\ref{fig:intprof}a shows the
average 2-30 keV pulse profile at 6 different intensities. Figure~\ref{fig:intprof}b shows the
peak-to-peak pulse fraction versus mean flux for each profile. In the following
descriptions and in Figures~\ref{fig:intprof}a and b, the pulse profiles are
numbered from one to six. (1) At the lowest intensities (6-17 \cpspcu, 2-30 keV), during the 2001 
outbursts, the profile's main feature was a deep notch, which we used to align 
the profiles. The profile consisted of an asymmetric structured main peak that 
peaked near phase 0.2. (2) As the intensity increased (23-65 \cpspcu), a second
notch becomes prominent near phase 0.45, making the profile consist of two main
peaks. These peaks were approximately equal in intensity at first, but the 
second peak brightens as the overall intensity increases. A small peak near 
phase 0.87 also appears. No {\em RXTE} PCA observations were taken at 
intensities between 65 and 178 \cpspcu. (3) The next profile shape occurs at intensities of 178-266 \cpspcu, near
the end of the 1998 outburst. This profile is markedly different from that near
the peak of the 2001 outburst (profile 2). Both notches are broader, the peak 
near phase 0.65 is brighter than the peak near phase 0.3 and the small peak at 
phase 0.87 has disappeared. The main notch appears to be less deep than at lower
intensities. (4) Profiles at intensities of 285-308 and 330 \cpspcu, all from 
the rise of the 1998 outburst, show a broader first peak and a less intense
shoulder following the second peak than the lower intensity profiles from the 
fall of the outburst (profile 3). (5) At intensities of 322-328 \cpspcu, during
the decline of the 1998 outburst, the first peak is narrower than at similar 
intensities during the outburst rise (profile 4), but broader than at lower 
intensities during the decline (profile 3). (6) At the peak of the outburst, the
width of the first peak is at an intermediate width between the profiles from 
the rise and fall of the outburst. During the 2001 outbursts, the profile shape
appeared to depend primarily on intensity, while during the 1998 bright 
outburst, the profile shape depended on both intensity and whether the profile 
was from the rise or fall of the outburst. 

\begin{figure}
\plotone{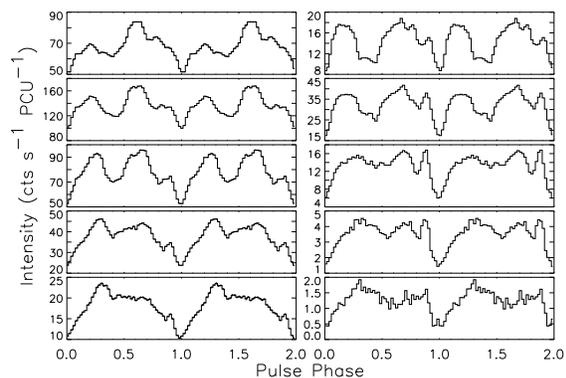}
\caption{Pulse profiles in 5 energy bands (from top to bottom) 2-5, 5-10,
10-15, 15-20, 20-30 keV from the peak of the 1998 outburst (left) and the 2001
outbursts (right).
\label{fig:profvsenergy}}
\end{figure}

Next we looked for variations in the pulse profile versus energy.
Figure~\ref{fig:profvsenergy} shows pulse profiles in 5 energy bands from the
peak of the 1998 outburst (left panel) and near one of the peaks in the pair of
outbursts in 2001 (right panel). In the 1998 profiles, the profile consists of
two main peaks at lower energies ($\lesssim 15$ keV). The first peak is dominant
at lower energies and the second peak becomes more dominant as energy increases.
Also, as energy increases, the notch between the two main peaks fills in. As the
2-30 keV intensity decreases, the profile ``shape" appears to move down in energy.
For example, in the right hand panel of Figure~\ref{fig:profvsenergy}, the 2-5
keV profile consists of two nearly equal peaks, reminiscent of the 10-15 keV
profile at the peak of the 1998 outburst. 

\subsection{Optical Observations}

A regular monitoring program was established to study the strength
and structure of the H$\alpha$ optical emission line during the X-ray
observations in 2001. This line arises from the circumstellar disk
surrounding the Be star and the line properties are strongly related
to the extent and dynamic structure of the disk. Data were primarily
collected from the 2.5m Isaac Newton Telescope (INT) in La Palma, Spain.
The INT was equipped with the 235-mm camera and EEV\#10 CCD. The use of 
the R1200R grating results in a nominal dispersion of 
$\approx0.40$\AA/pixel. Intermediate resolution spectroscopy was also performed
on 17th July 2001 and 22nd October 2001 using the 1.93-m telescope at the 
Observatoire de Haute Provence, France. That telescope was equipped with the 
long-slit spectrograph {\em Carelec} and the $1024\times2048$ EEV CCD. We used 
the 1200 ln/mm grating in first order, resulting in a nominal dispersion of 
$\approx 0.45$\AA/pixel. See Table~\ref{tab:ha} for details.

\begin{figure}
\plotone{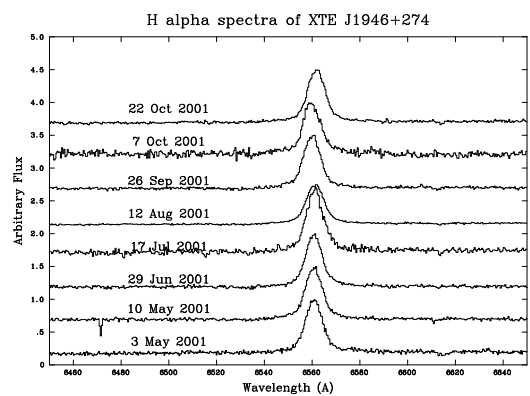}
\caption{H$\alpha$ profiles from 7 observations of XTE J1946+274 during our 2001
X-ray observations. The small dip near 6470 \AA\ in the 10 May 2001 spectrum is
most likely due to a cosmic ray hit.
\label{fig:ha}}
\end{figure}

The H$\alpha$ profile obtained from each observation is presented in
Figure~\ref{fig:ha}. The lines have been shifted vertically by an arbitrary
amount in order to present each profile clearly. The data of the 17 July and 7 
October 2001 are of a lower standard due to poor weather conditions. Nonetheless
it is possible to accurately determine the equivalent width of each
of the emission lines; values are presented in Table~\ref{tab:ha}. From 
these numbers one can see that there is little evidence for any changes in
the line flux, assuming a constant continuum. However, from Figure~\ref{fig:ha},
one can see obvious shifts in the line position after 2001 July.

\begin{figure}
\plotone{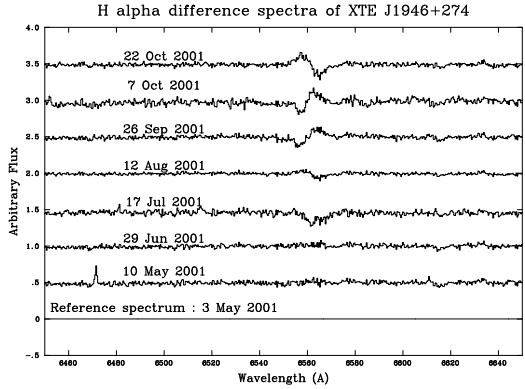}
\caption{H$\alpha$ profiles in Figure~\protect\ref{fig:ha} were normalized to
the same peak value and then the profile from 2001 May 3 was subtracted from all
other profiles to illustrate differences in the profiles. At some time between 
2001 Jun 29 and Jul 17, the profile began to change rapidly.
\label{fig:diff}}
\end{figure}

If one takes the first spectrum (3 May 2001) as a base line and
subtracts it from each of the other spectra after they have been
normalized to the same peak value, then significant changes are apparent. The
result of this process is shown in Figure~\ref{fig:diff}. It is apparent that
little happens to the profiles (and, presumably, the circumstellar
disk) until July/August 2001. Starting with the spectrum of 17 July 2001
one can see significant perturbations occurring in the line profile
indicative of density changes in the disk structure. Between August
and September the direction of the perturbation changes, suggesting
that the density enhancement/rarefication had changed sides (or
rotated around) in the disk. The size of the perturbation also seems
to have been increasing from July to October, though the separation of
the red and blue peaks remains approximately constant at $\sim$340 km/s.

\section{Discussion}

XTE J1946+274 is a very unusual Be/X-ray binary system and its behavior does not
fit well into the standard normal/giant outburst behavior. Instead it showed an
extended period of activity from 1998 September - 2001 July when the X-ray flux showed
considerable modulation, with two peaks or outbursts per orbit, but the X-ray flux
never dropped below the detection threshold of {\em RXTE}.  This extended period
of activity more closely resembles a series of normal outbursts than a single 
giant outburst; however, unlike typical normal outbursts, these outbursts shift
rapidly in orbital phase, there are two outbursts per orbit, and the X-ray flux
does not drop dramatically between outbursts. The initial outburst shows considerable
spin-up $\dot \nu \gtrsim 10^{-11}$ Hz s$^{-1}$, like a giant outburst, but this
spin-up is not substantially larger than that seen in later outbursts. In
addition, the peak flux (2-60 keV) of the initial outburst is about $7 \times 10^{-9}$
\ergcms, while the peak flux (2-60 keV) of the last two outbursts observed with
the PCA is about $2 \times 10^{-9}$ \ergcms, only a factor of $\sim 3.5$ fainter
than the brightest outburst. These fluxes correspond to luminosities of 
$8 \times 10^{37}$ \ergss and $2 \times 10^{37}$ \ergss, respectively, assuming 
our best-fit distance of 9.5 kpc.

In giant outbursts of Be/X-ray binaries, accretion disks are expected to be
present and indeed, evidence for an accretion disk, based on correlations 
between the observed flux and spin-up rate, has been found for several sources
during giant outbursts \citep{Parmar89, Reynolds96, Bildsten97, Wilson98, 
Stollberg99}. Independent evidence for an accretion disk based on the detection
of quasi-periodic oscillations during a giant outburst has been found for EXO 
2030+375 \citep{Angelini89} and A0535+262 \citep{Finger96b}. Until recently, 
normal outbursts were believed to be due to direct wind accretion from the Be 
disk, so significant spin-up was not expected because direct wind accretion is 
not believed to be very efficient at transferring angular momentum 
\citep{Ruffert97, Benensohn97, Ruffert99}. If enough angular momentum is present
in the accreted material, an accretion disk will form. However, evidence for 
spin-up during normal outbursts has been observed in GS 0834--430 
\citep{Wilson97}, 2S 1417--624 \citep{Finger96a}, 2S 1845--024 \citep{Finger99},
and in EXO 2030+375 \citep{Stollberg99, Wilson02}. 

We see a correlation between spin-up and flux for
XTE J1946+274 as shown in Figure~\ref{fig:fdotvsflux}. The spin-up rate and its
correlation with bolometric flux during XTE J1946+274's outbursts suggest an 
accretion disk may be present. In addition, our calculations show that the
specific angular momentum of the accreted material is comparable to the
Keplerian specific angular momentum, hence a disk is expected to form in XTE
J1946+274. Further, our fit to the spin-up vs. bolometric flux correlation, which
assumed a disk was present, yielded a distance of $9.5 \pm 2.9$ kpc which is 
consistent with the distance of 8-10 kpc derived from optical observations
\cite{Verrecchia02}. Hence an accretion disk is likely to be present in XTE
J1946+274.

Between 2001 July 31 (MJD 52121) and 2001 August 9 (MJD 52130), XTE J1946+274
dropped below the PCA's detection threshold. At approximately the same time,
between 2001 June 29 (MJD 52089) and 2001 July 17 (MJD 52107), the H$\alpha$
profile began to change rapidly. The profile was stable for 3 observations
on 2001 May 3, May 10, and June 29 (MJD 52032, 52039, \& 52089) that 
corresponded to X-ray observations during the decline of the second to last
outburst, during the low state between the last two outbursts, and near the peak
of the last outburst.  The coincidence of the change in H$\alpha$ with the X-ray
turn-off suggests that changes in the Be disk caused the X-ray outbursts to
cease.

To determine whether or not the observed X-ray turn-off was due to centrifugal
inhibition of accretion \citep{Stella86}, we estimate the flux at the onset of 
this effect by equating the magnetospheric radius to the corotation radius. The magnetospheric
radius is given by Equation~\ref{eqn:rm} and the corotation radius is given by
\begin{equation}
r_{\rm co} = (G M)^{1/3} (2 \pi \nu)^{-2/3}\label{eqn:rco}
\end{equation}
where $\nu$ is the spin frequency of the pulsar. Setting $r_{\rm m} = r_{\rm
co}$ gives the threshold flux for the onset of centrifugal inhibition of
accretion, i.e.,
\begin{equation}
F_{\rm x}^{\rm min} \simeq 3.0 \times 10^{-9}\ {\rm ergs}\ {\rm cm}^{-2}\
{\rm s}^{-1}\ k^{7/2} \mu_{30}^2 M_{1.4}^{-2/3} R_6^{-1} P_{\rm 15.8s}^{-7/3} 
d_{\rm kpc}^{-2}\label{eqn:cia}
\end{equation}
where $\mu_{30}$, $M_{1.4}$, $R_6$, and $P_{\rm 15.8 s}$ are the pulsar's 
magnetic moment in units of $10^{30}$ G cm$^{3}$, mass in units of 1.4 
$M_{\odot}$, radius in units of $10^6$ cm, and spin period in units of 15.8 
seconds, respectively. Using $d = 9.5$ kpc and $\mu = 1.6 \times 10^{30}$ G 
cm$^3$ yields $F_{\rm x}^{\rm min} \simeq 1.2 \times 10^{-10}$ \ergcms
$k^{7/2}$. For $k=0.47-0.91$, $F_{\rm x}^{\rm min} \simeq (0.6-6.0) \times
10^{-11}$ \ergcms. Our measured upper limit fluxes are in the range $(4-7)
\times 10^{-12}$ \ergcms, consistent with XTE J1946+274 entering the centrifugal
inhibition of accretion regime when the X-rays became undetectable with the PCA.

\section{Conclusions}

\begin{figure}
\plotone{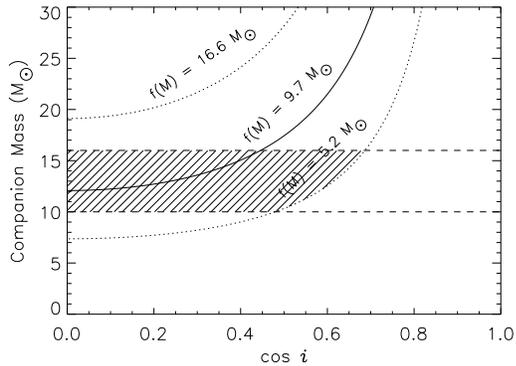}
\caption{Companion Mass versus the cosine of the inclination angle. The solid
line is the mass function from the fourth column of Table~\ref{tab:orb}. Dotted
lines denote the 68\% confidence error region on the mass function. Dashed lines
indicate the range of companion masses derived from optical observations. The
hatched region denotes the allowed masses and inclinations.
\label{fig:mc}}
\end{figure}

We propose that XTE J1946+274 is a system in which the Be star's equatorial
plane and the orbital plane are not aligned. This is suggested by: (1) The 
optical observations show a relatively narrow FWHM (8.6 \AA\ on 2001 May 10, which corresponds to $\pm$ 200 
km s$^{-1}$) for the single-peaked H$\alpha$ line, indicating that the
Be star is viewed from a relatively low inclination angle, i.e., nearly pole-on.
(2) The orbital signature is quite obvious in the pulse frequencies, indicating
that the orbital inclination angle is likely not low: the derived
mass function $f(M) = 9.7^{+6.9}_{-4.5} M_{\odot}$ from the piecewise frequency
model indicates an inclination angle $\gtrsim 46\arcdeg$\ for the expected mass
range of $10-16 M_{\odot}$ derived from optical observations. 
Figure~\ref{fig:mc} shows the range of allowed inclinations using the 1-$\sigma$
errors on the mass function.  

We see two outbursts per orbit in XTE J1946+274; however, these outbursts are 
not fixed in orbital phase. This combination of behaviors is unique to XTE 
J1946+274 and is likely not due to a single mechanism. The optical data and mass
function suggest that the orbital plane and the Be disk are not aligned in this
system. Such a misaligned system would be expected to produce two outbursts per
orbit, each approximately corresponding to the neutron star's passage through 
the Be disk; hence these outbursts would be expected to be fixed in orbital 
phase. For XTE J1946+274, we are faced with the difficult task of explaining not
only why we see the expected two outbursts per orbit, but also why these 
outbursts are not fixed in orbital phase. We propose that given the relatively 
high luminosities and spin-up rates in all of the outbursts that perhaps XTE 
J1946+274's unusual outburst behavior can be explained as a giant outburst. In 
normal outbursts, the Be disk is expected to be truncated at a resonance radius 
by tidal forces from the neutron star's orbit \citep{Okazaki01}; however, in giant outbursts, 
where much more material is believed to be present in the disk, it is believed 
that the disk is no longer truncated. Hence more material likely means a larger
disk around the Be star. Density perturbations propagating in the Be disk could
change the contact points of the disk and the orbit. These perturbations would 
have to be moving quite rapidly in the disk. This idea of a giant outburst 
combined with density perturbations seems most likely to work in a system where
the inclination angle between the Be disk and the orbital plane is fairly small.
It is not clear that this angle is small in XTE J1946+274. We do not claim that
the idea of a giant outburst in an inclined system fully explains the behavior 
observed from XTE J1946+274. Instead we put the idea forth as a suggestion to 
those doing simulations of these systems. 

\begin{figure}
\plotone{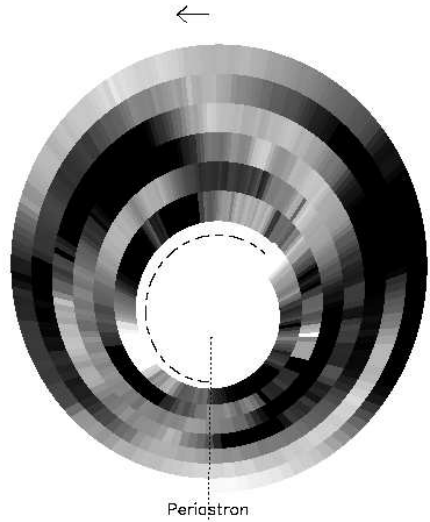}
\caption{Normalized intensity (darker colors indicate higher intensities) vs.
position in the orbit. The outside loop is the shape of the actual orbit. The
semi-major axis of each successive orbit is reduced by 10\% to offset the
orbits, allowing comparison of the orbital phase of the outburst in successive
orbits. This plot is not meant to indicate that the neutron star is spiraling in
toward its companion.
\label{fig:orbphase}}
\end{figure}

The phasing of the outbursts of XTE J1946+274 (see Figure~\ref{fig:orbphase}) is
difficult to understand. However, it is not the first Be/X-ray binary to show
shifts in outburst phase. EXO 2030+375 underwent a fairly sudden (within 4
orbits) shift in the peak phase of the outbursts from 6 days after periastron
(phase $\sim 0.13$) to 2.5 days before periastron (phase -0.05), followed by a
gradual shift in outburst phase to  2.5 days after periastron (phase 0.05). This
shift was believed to be associated with a density perturbation observed in the 
Be disk via H$\alpha$ observations \citep{Wilson02}. GS 0834--430 also underwent
sudden shifts in outburst phase, undergoing 9 outbursts approximately centered
on periastron, followed by dramatic shifts to outbursts centered on phases 0.37
and 0.75 \citep{Wilson97}. A companion was not known at the time for 
GS 0834--430, so we cannot confirm that density perturbations in the Be disk
were responsible for these phase shifts. In XTE J1946+274, we have a similar
problem, with the companion not observed for much of the time the X-ray
outbursts were occurring. In addition, the problem is compounded by the fact
that we are likely viewing the Be star nearly pole-on; hence the range of
projected rotational velocities is very small, making it difficult to detect
perturbed velocities which indicate density perturbations. Only large density
perturbations, such as the one that coincided with the X-ray turn-off, are
likely to be detected.

\acknowledgements
We are grateful to Luisa Morales for providing one of the optical spectra. This
research has made use of data obtained from the High Energy Astrophysics Science
Archive Research Center (HEASARC), provided by NASA's Goddard Space Flight 
Center (GSFC). {\em RXTE} ASM quick-look results were provided by the ASM/{\em 
RXTE} teams at MIT and at the GSFC SOF and GOF. The INT is operated on the 
island of La Palma by the Isaac Newton Group in the Spanish Observatorio del 
Roque de los Muchachos of the Instituto de Astrof\'{\i}sica de Canarias. Based 
in part on observations made at Observatoire de Haute Provence (CNRS), France.

\begin{deluxetable}{lllll}
\tabletypesize{\scriptsize}
\tablecaption{Observations of the Sky Region Including XTE J1946+274}
\tablewidth{0pt}
\tablehead{\colhead{Dates} & \colhead{MJD} & \colhead{Instrument} &
\colhead{No. Obs.\tablenotemark{a}} & 
 \colhead{Total PCA Exposure (ks)\tablenotemark{a}}}
\startdata
1991 Apr 15 - 2000 May 27 & 48361-51691 & BATSE & \ldots & \ldots \\
1996 Feb 23 - 2002 Apr 11 & 50136-52376 & {\em RXTE} ASM & \ldots & \ldots \\
1998 Sep 16 - 1998 Oct 14 & 51072-51100 & {\em RXTE} PCA, HEXTE & 12 & 85.87 \\
2001 Mar 9 - 2001 Sep 25 &  51977-52177 & {\em RXTE} PCA, HEXTE & 24 & 129.49 \\
\enddata
\tablenotetext{a}{For pointed observations.}
\label{tab:obs}
\end{deluxetable}

\begin{deluxetable}{llll}
\tabletypesize{\scriptsize}
\tablecaption{Orbit Fit}
\tablewidth{0pt}
\tablehead{\colhead{Parameter} & \colhead{Linear $\nu_{\rm emit}$} &
\colhead{10th Order $\nu_{\rm emit}$} & \colhead{Piecewise $\nu_{\rm emit}$}} 
\startdata
$P_{\rm orb}$ (days)& $173 \pm 2$ & $167.8 \pm 0.6$ & $169.2 \pm 0.9$ \\
$\tau_{\rm p}$ & JD$2451524.6 \pm 16.9$ & JD$2451572.3 \pm 1.6$ & JD$2451558.7 \pm 4$ \\
$a_{\rm X} \sin i$ (lt-sec) & $972 \pm 118$ & $474 \pm 14$ & $640 \pm 120$ \\
$e$ & $0.19 \pm 0.13$ & $0.35 \pm 0.03$ & $0.33 \pm 0.05$ \\
$\omega$ (degrees) & $-153 \pm 37$ & $-38 \pm 5$ & $-91 \pm 23$ \\
$f(M)$ $(M_{\odot})$ & $32 \pm 11$ & $4.0 \pm 0.3$ & $9.7^{+6.9}_{-4.5}$ \\
$\chi^2/$d.o.f & $2699556/101$ & $28756/92$ & $220.1/37$ \\
\enddata
\label{tab:orb}
\end{deluxetable}

\begin{deluxetable}{cccccc}
\tabletypesize{\scriptsize}
\tablecaption{Journal of optical observations.}
\tablewidth{0pt}
\tablehead{\colhead{Date} & \colhead{Telescope\tablenotemark{a}} &
\colhead{Instrument\tablenotemark{b}}&
\colhead{Grating}&\colhead{Detector}&\colhead{H$\alpha$ EW (\AA)}}
\startdata
 3 May 2001&INT&IDS+500 camera&R1200R  &TEK5 &45.9$\pm$0.6\\
10 May 2001&INT&IDS+235 camera&R1200R  &EEV10&41.0$\pm$0.8\\
29 Jun 2001&INT&IDS+235 camera&R1200R  &EEV10&41.1$\pm$0.3\\
17 Jul 2001&OHP&Carelec       &1200l/mm&EEV  &39.1$\pm$0.4\\
12 Aug 2001&INT&IDS+235 camera&R1200R  &EEV10&43.9$\pm$0.3\\
26 Sep 2001&INT&IDS+235 camera&R1200R  &EEV10&43.6$\pm$0.3\\
 7 Oct 2001&INT&IDS+235 camera&R1200R  &EEV10&38.5$\pm$0.5\\
22 Oct 2001&OHP&Carelec       &1200l/mm&EEV  &42.5$\pm$0.3\\
\enddata
\tablenotetext{a}{INT = Isaac Newton Telescope, OHP = Observatoire de Haute
Provence}
\tablenotetext{b}{IDS = Intermediate Dispersion Spectrograph.}
\label{tab:ha}
\end{deluxetable}



\end{document}